\documentclass[9pt]{article}

\usepackage{amsmath}
\begin{document}

\title{\bf A Discussion on the Different Notions\\ of Symmetry  of 
Differential Equations}

\author{GIAMPAOLO CICOGNA \\ ~ \\
{\it Dipartimento di Fisica ``E.Fermi'' dell'Universit\`a di Pisa} \\
{\it and 
  Istituto Nazionale di Fisica Nucleare, Sez. di Pisa} \\ 
 {\it Largo B. 
Pontecorvo 3, Ed. B-C, I-56127, Pisa, Italy} \\
(fax: +39-050-2214887; e-mail: cicogna@df.unipi.it)}

\newtheorem{theorem}{Theorem}
\newtheorem{lemma}{Lemma}
\newtheorem{proposition}{Proposition}

\newtheorem{definition}{Definition}

\def \ov{\over}
\def \bar{\overline}
\def \beq{\begin{equation} }
\def \eeq{\end{equation} }
\def \lb{\label}

\def \pd{\partial}
\def\~#1{\widetilde #1}

\def \sy {symmetry}
\def \sys {symmetries}
\def \so {solution}
\def \eq{equation}
\def \R{{\bf R}}

\def\a{\alpha}
\def\be{\beta}
\def\phi{\varphi}
\def\De{\Delta}
\def\la{\lambda}
\def\ga{\gamma}

\def\Na2{\nabla^2}

\def \qq{\qquad}
\def \q{\quad} 
\def\={\, =\, }
\def\vf{vector field }
\baselineskip .5cm
\date{}
\maketitle
\def\sk{\smallskip}

\noindent
\def \ov{\over}
\def \bar{\overline}
\def \nn{\nonumber}

\def \pd{\partial}
\def\~#1{\widetilde #1}
\def\.#1{\dot #1}
\def\^#1{\widehat #1}
\def\d{{\rm d}}       
\def \id{\! :=}

\def \sy {symmetry}
\def \sys {symmetries}
\def \so {solution}
\def \eq{equation}
\def \R{{\bf R}}

\def\a{\alpha}
\def\be{\beta}
\def\phi{\varphi}
\def\de{\delta}
\def\g{\gamma}
\def\De{\Delta}
\def\th{\theta}
\def\ka{\kappa}
\def\s{\sigma}

\def \qq{\qquad}
\def \q{\quad}
\def \pn{\paragraph\noindent}
\def \sk{\medskip}
\def \noi{\noindent}
\def\BE{Boussinesq equation}
\def\CS{conditional symmetry}
\def\={\, =\, }
\begin{abstract}
A discussion is presented, within a simple unifying scheme,
about different types of symmetry of PDE's, with the introduction and a
precise characterization of the notions of ``standard'' and  ``weak'' 
conditional symmetries, together with their relationship with exact 
and partial symmetries.  An extensive use of ``symmetry-adapted''
variables will be made, and  some clarifying examples are provided.
\end{abstract}
\section{Introduction}

This paper is essentially a presentation of a unifying and comprehensive 
scheme, where several different notions of \sy\ for differential problems 
may be considered and compared. In particular, this approach will permit 
the introduction of  ``subtler" notions of conditional \sys\ (or 
``nonclassical \sys '')
\cite{cicogna:BC,cicogna:BC1,cicogna:FK,cicogna:Fu,cicogna:W2}, with a
clear  distinction and characterization of these \sys\ and of other
related  concepts, including the more recently introduced notions of
``partial \sys '' \cite{cicogna:CG} 
(see also \cite{cicogna:C,cicogna:C2}), and of ``hidden \sys " 
(see e.g. \cite{cicogna:abr,cicogna:iri}). 

For simplicity, we will consider here only the case of partial 
differential \eq s (PDE) 
\begin{equation} \De\equiv \De_a(x,u^{(m)})\=0 \qq\qq (a=1,\ldots , \nu)
\label{cicogna:De1} \end{equation}
for the $q$ functions $u_\a=u_\a(x)$ of the $p$ variables $x_i$
(as usual, $u^{(m)}$ denotes the functions $u_\a$ together with their
$x$ derivatives up  to the order $m$), and only ``geometrical" or 
Lie-point \sys\ (i.e., no  generalized or B\"acklund, potential or 
nonlocal, and so on), i.e. \sys\ generated by vector fields of the form 
(sum over repeated  indices)
\begin{equation} X\= \xi_i(x,u){\pd\over {\pd x_i}}+
\phi_\a(x,u){\pd \over{\pd u_\a}} \label{cicogna:X1}
\end{equation}
although the relevant results could be suitably extended also to
these \sys , whose importance is well known and also recently further
emphasized (cf. e.g. \cite{cicogna:GMR,cicogna:FF,cicogna:Zh,
cicogna:Sop}).

\section{Exact Symmetries}

Let us start with the basic and standard definition, with the usual 
``blanket" assumptions tacitly understood (see
\cite{cicogna:Ov,cicogna:Ol,cicogna:Ib,cicogna:Ga,cicogna:BA}):
\begin{definition}
A system of  PDE $\De_a(x,u^{(m)})\=0$
is said to admit the Lie-point \sy\ generated by the vector field $X$
(or to be {\em symmetric} under $X$) if the following condition
\begin{equation} X^*(\De)|_{\De=0}\=0 \label{cicogna:XXDD} \end{equation}
is satisfied, or -- equivalently (at least under mild hypotheses)
-- if there are functions $G=G_{ab}(x,u^{(m)})$ such that
\begin{equation} (X^*(\De))_a\=G_{ab}\,\De_b \ .\label{cicogna:XDgD}
\end{equation}
\end{definition}

\sk\noi
We simply denote by $X^*$ the ``appropriate'' prolongation of $X$ for 
the \eq\ at hand, or -- alternatively -- its infinite prolongation 
(indeed, only a finite number of terms will appear in calculations).

\sk\noi
Let us also give this other definition:
\begin{definition}
A system of PDE as before is said to be {\em invariant} under 
$X$  if
\begin{equation} X^*(\De)\=0 \ . \label{cicogna:inv}\end{equation}
\end{definition}

\sk
For instance, the Laplace \eq\ $u_{xx}+u_{yy}=0$ is {\em invariant}
under the rotation \sy\  $X=y\pd/\pd x-x\pd/\pd y$; the heat \eq\
$u_t=u_{xx}$ is {\em symmetric} but {\em not} invariant under
\[ X=2t{\pd\ov \pd x}-xu{\pd\ov \pd u} \]
indeed one has
$\, X^*(u_t-u_{xx})= -x (u_t-u_{xx})$.

\sk
We then have:

\begin{theorem}
Let $\De=0$ be a nondegenerate system of PDE's, symmetric under a 
projectable vector field $X$, according to Def.~1. Then, there are  
new $p+q$ variables $s,z$ and $v$, with $s\in\R$, $z\in \R^{p-1}$ and 
$v\equiv(v_1(s,z),\ldots,v_q(s,z))$, and a new system of PDE's, say
$K=0$, with $\q K_a\=S_{ab}(s,z,v^{(m)})\,\~\De_b(s,z,v^{(m)})$ (where
$v^{(m)}$  stands for $v(s,z)$ and its derivatives with respect to 
$s$ and $z$, and $\~\De=\~\De(s,z,v^{(m)})$
is $\De$ when expressed in terms of the new variables $s,z,v$),
which is locally {\rm equivalent} to the initial system and is {\rm
invariant} (as in Def.~2) under the \sy\ $X=\pd/\pd s$, i.e.
$K_a=K_a(z,v^{(m)})$.
\end{theorem}
{\tt Proof} (a sketch).
Given the \sy\ $X$, one has to introduce ``canonical variables'' (or
\sy -adapted variables) $s,z\equiv(z_1,\ldots,z_{p-1})$, which are 
defined by
\[
X\ s\,\equiv\,\xi_i {\pd s\over{\pd x_i}}+\phi_\a {\pd s\over{\pd
u_\a}}\=1 \q ; \q X\ z_k \= 0 \q (k=1,\ldots, p-1)
\]
Using the method of characteristics, one also finds the $q$  
dependent variables $v=v_\a(s,z)$; once
written in these coordinates, the \sy\ field and all its prolongations
are simply given by
\begin{equation} \~X=\~X^*={\pd\ov \pd s} \label{cicogna:XX*}\end{equation}
whereas the \sy\ condition (Def. 1) becomes 
$\, 
\displaystyle{{\pd\~\De\over{\pd s}}\Big|_{\~\De=0}\= 0}\, $, or
\begin{equation} {\pd\over{\pd s}}\~\De_a\=G_{ab}\~\De_b
\label{cicogna:dsDgD} \end{equation}
It is not difficult to show (cf. \cite{cicogna:CL}) that for any $\~\De_a$ 
satisfying the system (\ref{cicogna:dsDgD})
there are smooth locally invertible functions $S_{ab}$ such 
that the combinations $K_a\id S_{ab}\~\De_b$ are independent of $s$, as 
claimed. We have assumed here
for convenience that the vector fields $X$ are ``projectable'',
or -- more explicitly -- that the functions $\xi$ in (\ref{cicogna:X1}) 
do not depend on $u$  (as happens in most cases in the study of PDE's)
in order to  simplify calculations  in the introduction of the canonical 
coordinates, and to get a more direct relationship between 
\sys\ and  symmetry-invariant \so s (for a discussion on this  point, cf.
\cite{cicogna:Pu}). $\hfill\triangle$

\medskip
It should be emphasized that the result in Theorem 1 is
{\it not} the same as (but includes) the well known result concerning
the reduction of the given \eq s to $X-$invariant \eq s for the invariant
variables $w(z)$: indeed, introducing the new ``\sy-adapted'' variables 
$s,z$ and $v(s,z)$, we have transformed the \eq s into  {\em equivalent} 
\eq s for $v(s,z)$. If one now {\em further} assumes
that $\pd v/\pd s=0$, i.e. if one looks for the $X-$invariant \so s
where $v=w(z)$, then the \eq s $K_a=0$ become a system of \eq s
\begin{equation} K_a^{(0)}(z,w^{(m)})=0 \label{cicogna:K0}\end{equation}
involving only the variables $z$ and functions depending only on $z$
(see \cite{cicogna:ZTP}
for a  general discussion on the reduction procedure).

\section{``Standard'' and ``weak'' conditional\\ symmetries, and related
notions}

Let us now consider the case of non-exact \sys . A fundamental and 
largely comprehensive notion has been introduced by Fushchych 
\cite{cicogna:FQ}: let us say that $X$ is a {\em \CS\ of the 
\eq\ $\De=0$ in the sense of Fushchych} if there is a supplementary 
\eq\ $E=0$ such that $X$ is an exact \sy\ of the system $\De=E=0$. 

The simplest and more common case is obtained choosing as supplementary
\eq\ the ``side condition'' or ``invariant surface condition" 
\begin{equation}  X_Q u\equiv \xi_i {\pd u\ov \pd {x_i}}-\phi\,
=\, 0 \label{cicogna:gcs} \end{equation}
where $X_Q$ is the \sy\ written in ``evolutionary form'' 
\cite{cicogna:Ol}: this corresponds to the usual (properly called)
\CS\ (CS)  (also called $Q-$conditional \sy ) and the above condition
indicates that we are looking precisely for \so s which are {\it
invariant} under~$X$.

To avoid unessential complications with notations, we will consider
from now on only the case of a single PDE $\De=0$ for a single unknown 
function $u(x)$. The extension to more general cases is in principle 
completely straightforward.

It is known that the above definition of CS suffers from some intrinsic 
difficulties, 
essentially due to the necessity of introducing and dealing with the 
differential consequences of (\ref{cicogna:gcs}) (for a discussion of this
point, see e.g. \cite{cicogna:Ol,cicogna:PS},
and \cite{cicogna:FT,cicogna:ZT,cicogna:Pop} for a more complete
definition). Related to these  difficulties is the quite embarrassing
sentence by Olver and  Rosenau \cite{cicogna:OR} (see also
\cite{cicogna:PS}), which says --  essentially -- that any vector 
field $X$ is a CS, and any \so\ of the given \eq\ is an invariant \so\ 
under some $X$.

To clarify this point, we will introduce a subtler definition of CS. This 
will be made resorting once again to the canonical coordinates 
$s,z,v=v(s,z)$, introduced in the proof of Theorem 1. First of all, in 
these coordinates the invariance condition $X_Qv=0$ becomes
\begin{equation} {\pd v\ov \pd s}\=0 \label{cicogna:us}\end{equation}
and the condition of CS  takes the 
simple form (let us now retain for simplicity the same notation $\De$, 
instead of $\~\De$, also in the new coordinates)
\begin{equation} {\pd \De\over {\pd s}}\Big|_{\Sigma}\=0 
\label{cicogna:DsS}\end{equation}
here $\Sigma$ stands for the set of the simultaneous \so s of $\De=0$
and   $v_s=\pd v/\pd s=0$, together with the derivatives of $v_s$
with respect to all the variables $s$ and $z_k$. Introducing  the
global notation $v_s^{(\ell)}$ to indicate $v_s,v_{ss},v_{sz_k}$ etc.,
we shall say that $X=\pd/\pd s$ is a CS in {\em standard 
sense}  if the \eq\ takes the form
\begin{equation} \De\=R(s,z,v^{(m)}) 
K(z,v^{(m)})+\sum_\ell\Theta_\ell(s,z,v^{(m)})\,
v_s^{(\ell)}\= 0 \label{cicogna:DRT}\end{equation}
where the point to be emphasized is that $K$ does not depend
{\em explicitly} on $s$, and $R,K$ do not contain $v_s^{(\ell)}$. 
It is then clear that, if one now looks for \so s of $\De=0$ which are 
independent on $s$, i.e.
such that $v_s^{(\ell)}=0$, or of the form $v=w(z)$,
then eq. (\ref{cicogna:DRT}) becomes a ``reduced'' \eq\ 
$K^{(0)}(z,w^{(m)})=0$, just as in the exact \sy\ case.

But this is clearly only a special case. Indeed, the \eq\ $\De=0$ may 
also take the form
\begin{equation} \label{cicogna:sr} \De\=\,\sum_{r=1}^\s s^{r-1} 
K_r(z,v^{(m)})+
\sum_\ell\Theta_\ell(s,z,v^{(m)})\, v_s^{(\ell)} \=0 \end{equation}
where the part not containing $v_s^{(\ell)}$ is a polynomial in the 
variable $s$, with coefficients $K_r$ not depending explicitly on $s$, or 
also -- more in general (with some different regrouping of the terms 
containing $s$ into linearly and functionally independent terms $R_r$)
\begin{equation} \De\=\sum_{r=1}^\s R_r(s,z,v^{(m)})\, K_r(z,v^{(m)})+
\sum_\ell\Theta_\ell(s,z,v^{(m)})\, v_s^{(\ell)}\=0  \label{cicogna:Kr}
\end{equation}
In this case, if one looks for $X-$invariant \so s
$w(z)$ of $\De=0$, one is faced with the system of reduced \eq s (not
containing $s$ nor functions of $s$)
\begin{equation} K_r^{(0)}(z,w^{(m)})\=0\qq ; \qq r=1,\ldots,\s 
\label{cicogna:Krz}\end{equation}
Assume that this system admits some \so\ (it is known that the 
existence of invariant \so s is by no means guaranteed in 
general, never for ``standard'' CS, nor for ``exact'' Lie \sys ), 
we will say that $X$ is {\em  weak CS of order $\s$}.

We now see that the set of the \so s of the above system  can be 
characterized equivalently as the set of the \so s of the system
\begin{equation}  \De\=0 \q ;\q {\pd \De\ov \pd s}\=0\q ;\,\ldots\, ; \q
{\pd^{\s-1}\De\ov \pd s^{\s-1}}\=0 \q ; \q v_s^{(\ell)}\=0 
\label{cicogna:d1s}\end{equation}

Coming back to the original coordinates $x,u$, the set of conditions
(\ref{cicogna:d1s}) becomes 
\begin{equation}  \De\=\De^{(1)}\=\ldots\=\De^{(\s -1)}\=0 \q ; 
\q X_Q\,u\=0 \label{cicogna:dddd}\end{equation}
where
\begin{equation} \De^{(1)}\id X^*(\De),\qq
\De^{(2)}\id X^*(\De^{(1)}) ,\qq \ldots
\label{cicogna:dd2}\end{equation}
(as already pointed out, also the differential consequences of 
$X_Qu=0$ must be taken into account), and a CS of order $\s$ can be 
characterized by the condition
\begin{equation} \label{cicogna:xss} X^*(\De)|_{\Sigma_\s}\= 0
\end{equation}
where $\Sigma_\sigma$ is the set (if not empty, of course) of the
\so s of the system (\ref{cicogna:dddd}).

\sk
We can summarize our discussion in the following form.

\begin{proposition} Given a PDE $\De=0$, a projectable vector field $X$ is 
a ``standard'' conditional \sy\
for the \eq\ if it is  a \sy\ for the system
\[   \De\=0\q ;\q  X_Qu\=0 \]
and this corresponds to the existence of a reduced \eq\ in $p-1$ 
independent variables, which -- if admits \so s --
gives   $X-$invariant \so s of  $\De=0$. A vector field $X$ is a
``weak'' CS (of order $\s$) if it is a \sy\ of the system
\[ 
 \De\=0 \  ;\  \De^{(1)}:=X^*(\De)\=0\ ;\   
\De^{(2)} := X^*(\De^{(1)})\=0\  ;\ \ldots\ ;\]
\[\De^{(\s-1)}=0\ ; \ X_Qu\=0
\]
and this corresponds to the existence of a system of $\s$
reduced \eq s, which --  if admits \so s -- gives  $X-$invariant  \so s 
of $\De=0$.  Introducing $X-$adapted variables $s,z$, such that $Xs=1,\,
Xz=0$, the  PDE has the form (\ref{cicogna:DRT}) in the case of standard 
CS, 
or (\ref{cicogna:Kr}) in the case  of weak CS.
\end{proposition}

\sk
If one neglects the invariance condition $X_Qu=0$, one is actually 
dealing with the case of {\em partial} \sys . Indeed (see 
\cite{cicogna:CG,cicogna:C,cicogna:C2}), $X$ is precisely a partial 
\sy\ of order $\s$ if $X$  is a \sy\ of the system
\begin{equation} \label{cicogna:dps}
\De\=\De^{(1)}\=\ldots\=\De^{(\s -1)}\=0 \end{equation}
If this is the case, $X$ maps one into another the \so s of the system 
(\ref{cicogna:dps}), which is then a ``symmetric set of \so s of
$\De=0$" \cite{cicogna:C}.  In particular, if in this set 
there are some \so s which are left fixed by $X$, then $X$ is also a CS
(either standard or weak) of $\De=0$.

We can then rephrase the Olver-Rosenau statement \cite{cicogna:OR} in 
the form:

\begin{proposition}
Any vector field $X$ is either an exact, or a standard CS, or a weak CS. 
Similarly, any $X$ is either an exact or a partial \sy . 
\end{proposition}

It is well known that the set of the \so s which can be  obtained 
in this way may be empty or contain only trivial \so s (e.g., $u=$ const): 
it is clear that the choice of good candidates as these ``non-exact'' \sy\ 
generators should be guided by some reasonable criterion and motivated 
guess. 

It is also clear that all the notions of non-exact \sys\ considered above 
can be viewed as special cases of CS in the sense of Fushchych. In all 
the above discussion, we have considered the case of a single vector 
field $X$; clearly, the situation becomes richer and richer if more than 
one vector field is taken into consideration. First of all, the reduction 
procedure itself must be adapted and refined when the given \eq\ admits 
an algebra of \sys\ of dimension larger than $1$ (possibly infinite): for 
a recent discussion see \cite{cicogna:GTW}. Secondly, for instance, it
can  happen that the reduced \eq s (\ref{cicogna:K0}) or
(\ref{cicogna:Krz})  may admit some new \sy\ $Y$ not shared by the
original \eq\ $\De=0$: this  is (essentially) the case of ``hidden \sys ''
\cite{cicogna:abr,cicogna:iri}. Different reduction procedures have been
also proposed, based on the  introduction of multiple  suitable 
differential constraints: see, e.g.,
\cite{cicogna:GMR,cicogna:Vo,cicogna:Nhe,cicogna:Kap}, and also 
\cite{cicogna:Ib}.

\sk

\section{Examples} 
We will give here some simple examples, to illustrate the properties 
of the different types of \sys\ introduced above, and the 
different  \so s that can be obtained accordingly.

\sk\noi
{\tt Example}. Consider the \eq , proposed by Popovich 
\cite{cicogna:Pop}
\[ u_t+u_{xx}-u+t(u_x-u)\=0 \qq ; \qq u=u(x,t) \]
The vector field $X=\pd/\pd t$ is not an exact nor a 
standard CS, but  is a weak CS, indeed the system of \eq s
(\ref{cicogna:Krz}) (here $s=t$) becomes  $u_{xx}\=u,\, u_x\=u$, with
\so\ $u=c\, \exp(x)$.
The same vector field $X=\pd/\pd t$ is a weak CS also for this  
variation of the above \eq :
\[  u_t-u_{tt}+u_{xx}-u+t(u_x-u)\=0 \]
with the same \so\  as above.
But $X$ is now also a partial \sy : indeed, the \eq\
$\De^{(1)}=0$ is now $u_x-u=0$, and combining it with $\De=0$  we find the 
more general \so\
$ u\=c \,\exp (x)+c_1\,\exp(x+t). $ 
Considering this other variation of the Popovich  example
\[  t^2(u_t-u)+u_{xx}-u+t(u_x-u)\=0 \]
here $X=\pd/\pd t$ is only a partial \sy , leading to the \so\ 
$u=c\,\exp(x+t)$  (strictly speaking, it is also a weak CS, but producing
only the trivial \so\ $u=0$ !).

\sk\noi{\tt Example.} It is well known that the  Korteweg - de Vries \eq
\[  u_t+u_{xxx}+uu_x=0 \qq ; \qq u=u(x,t) \]
does not admit (standard) CS, apart from its exact \sys . 
There are however weak CS; e.g. the scaling
\[ X\=2x{\pd\ov{\pd x}}+t{\pd\ov{\pd t}}+u{\pd\ov {\pd u}} \]
is an exact \sy\ for the system 
$\, \De=0,\,\De^{(1)}=X^*(\De)=0,\, X_Qu=0$,
which means that this is a weak CS, giving the scaling-invariant \so\ 
$u=x/t$.
But also, if we neglect the invariance condition
$X_Qu=0$, we obtain the (clearly larger) symmetric set of \so s 
$\, u=(x+c_1)/(t+c_2) $,
showing that the above $X$ is also a partial \sy .

\sk\noi{\tt Example.}
The \sy\ properties of the \BE\ 
\begin{equation} \label{cicogna:BE}
u_{tt}+u_{xxxx}+u\, u_{xx}+u_x^2\= 0\qq ; \qq u=u(x,t)
\end{equation}
have been the object of several papers (see e.g. 
\cite{cicogna:GMR,cicogna:LW,cicogna:Cl}).
For what concerns standard CS, writing the general vector field in
the form
\begin{equation} X\, =\, \xi(x,t,u){\pd\over {\pd x}}+\tau(x,t,u){\pd 
\over{\pd
t}}+\phi(x,t,u){\pd \over{\pd u}} \label{cicogna:gsy}
\end{equation}
a complete list of CS has been given both for the case 
$\tau\not=0$ (and therefore, without any restriction, 
$\tau=1$)  \cite{cicogna:LW} and for the case $\tau=0$
\cite{cicogna:Cl,cicogna:Lo}; it has been also shown that the 
invariant \so s under these CS are precisely those found  
by means of the ``direct method", which is not based on (but clearly 
related to) \sy\ properties \cite{cicogna:LW,
cicogna:Cl,cicogna:CK}.

To complete the analysis, one can also look for  \sys\ with $\xi=0$. It 
is not difficult to verify   that no standard CS of this form is 
admitted.  There are however  {\em weak} CS: an example is
\begin{equation}\label{cicogna:X1t} X\={\pd\ov{\pd 
t}}+\Big({1\ov{t^2}}-{2u\ov{t}}\Big){\pd
\ov{\pd u}}\end{equation}
one obtains from this $s=t,z=x$ and  $u(x,t)=t^{-1}+t^{-2}v(x,t)$,
giving 
\begin{equation} v v_{x x} + v_x^2 +6v + t(v_{xx}+2) + t^2 v_{xxxx}- 4 t 
v_t + t^2 v_{tt}  \= 0
\label{cicogna:BExt}\end{equation}
which is precisely of the form (\ref{cicogna:Kr}) (the role of $s$ is 
played here by $t$). Looking indeed for \so s with $v=w(x)$, one gets  
a system of the three ODE's 
\[v v_{x x} + v_x^2 +6v=0,\q v_{xx}+2=0,\q 
v_{xxxx}=0 \]
(cf. (\ref{cicogna:d1s})), admitting the common \so\ $w=-x^2$ and giving
the  (quite elementary) \so\ $u=1/t-x^2/t^2$ of the \BE . 

Another example of weak CS for the \BE\ is the following
\begin{equation} \label{cicogna:CSP} X\= t^2{\pd\ov \pd x}+{\pd\ov \pd 
t}-\Big(2x+{10\ov
3}t^3\Big){\pd \ov \pd u}\end{equation} 
now $s=t,z=x-t^3/3$ and $u=-2sz-s^4+v(s,z)$.
The additional \eq s $\De^{(1)}=X^*(\De)=0$ 
etc.: now become
\begin{equation} \De^{(1)}\= -10  t - 3 u_x - 2 t u_{xt} - 
              {5\ov 3} t^3 u_{x  x} - x u_{x  x} \= 0 
\q ; \q \De^{(2)}\= 
2 + u_{x t} + t^2 u_{x x} \= 0 \label{cicogna:CP2}\end{equation}
and taking into account also the invariance condition $X_Qu=0$, we easily 
conclude that this is a weak CS of order $\s=3$ and obtain the \so\ 
\begin{equation} u(x,t)\= -{t^4\ov 3} - 2t x - {12\ov (x - t^3/3)^2} 
\label{cicogna:CP3}\end{equation}
If instead we do {\em not} impose the invariance condition
$X_Qu=0$ and solve the three \eq s 
(\ref{cicogna:BE},\ref{cicogna:CP2}), we find,
in  addition to the invariant \so\ (\ref{cicogna:CP3}), also the following 
family of \so s
$ u(x,t)\= - t^4/3 + c_1 t - 2t x + c_2 $, showing that the above \sy\ is 
also a partial \sy .



\end{document}